\documentclass[]{spie}  %>>> use for US letter paper
\usepackage{amsmath,amsfonts,amssymb}
\usepackage{graphicx}
\usepackage[colorlinks=true, allcolors=blue]{hyperref}
\usepackage{aas_macros}

\title{High-contrast imager for complex aperture telescopes (HiCAT): 8. Dark zone demonstration with simultaneous closed-loop low-order wavefront sensing and control}

\author[a]{R\'emi Soummer}
\author[a]{Emiel H. Por}
\author[b]{Raphaël Pourcelot}
\author[c]{Susan Redmond}
\author[d]{Iva Laginja}
\author[e]{Scott D. Will}
\author[a]{Marshall D. Perrin}
\author[a]{Laurent Pueyo}
\author[a]{Ananya Sahoo}
\author[a,f]{Peter Petrone}
\author[a]{Keira J. Brooks}
\author[a]{Rachel Fox}
\author[a]{Alex Klein}
\author[a]{Bryony Nickson}
\author[a]{Thomas Comeau}
\author[g]{Marc Ferrari}
\author[a]{Rob Gontrum}
\author[h]{John Hagopian}
\author[d]{Lucie Leboulleux}
\author[a]{Dan Leongomez}
\author[a]{Joe Lugten}
\author[i]{Laurent M. Mugnier}
\author[b]{Mamadou N'Diaye}
\author[a]{Meiji Nguyen}
\author[a]{James Noss}
\author[g,i]{Jean-François Sauvage}
\author[j]{Nathan Scott}
\author[a]{Anand Sivaramakrishnan}
\author[e]{Hari B. Subedi}
\author[a]{Sam Weinstock}
\affil[a]{Space Telescope Science Institute, 3700 San Martin Drive, Baltimore, USA}
\affil[b]{Université Côte d’Azur, Observatoire de la Côte d’Azur, CNRS, Laboratoire Lagrange, France }
\affil[c]{Dept. of Mechanical and Aerospace Engineering, Princeton University, Princeton, NJ, USA}
\affil[d]{LESIA, Observatoire de Paris, Université PSL, Sorbonne Université, Université Paris Cité, CNRS, 5 place Jules Janssen, 92195 Meudon, France}
\affil[e]{NASA Goddard Space Flight Center, Greenbelt, MD 20771, USA}
\affil[f]{Hexagon Federal, Chantilly, VA 20151, USA}
\affil[g]{Aix Marseille Université, CNRS, CNES, LAM (Laboratoire d’Astrophysique de Marseille) UMR 7326, 13388 Marseille, France}
\affil[h]{Advanced Nanophotonics Inc., 4437 Windsor Farm Rd,
Harwood, MD USA}
\affil[i]{DOTA, ONERA, Universit\'e Paris Saclay, F-92322 Ch\^{a}tillon, France}
\affil[j]{John Hopkins University, 3400 North Charles Street, Baltimore, MD 21218, USA}

\authorinfo{Further author information, send correspondence to R.S.: E-mail: soummer@stsci.edu}

\begin{document} 
\maketitle

\begin{abstract}
We present recent laboratory results demonstrating high-contrast coronagraphy for the future space-based large IR/Optical/Ultraviolet telescope recommended by the Decadal Survey. The High-contrast Imager for Complex Aperture Telescopes (HiCAT) testbed aims to implement a system-level hardware demonstration for segmented aperture coronagraphs with wavefront control. The telescope hardware simulator employs a segmented deformable mirror with 37 hexagonal segments that can be controlled in piston, tip, and tilt. In addition, two continuous deformable mirrors are used for high-order wavefront sensing and control. The low-order sensing subsystem includes a dedicated tip-tilt stage, a coronagraphic target acquisition camera, and a Zernike wavefront sensor that is used to measure and correct low-order aberration drifts.  We explore the performance of a segmented aperture coronagraph both in “static” operations (limited by natural drifts and instabilities) and in “dynamic” operations (in the presence of artificial wavefront drifts added to the deformable mirrors),  and discuss the estimation and control strategies used to reach and maintain the dark-zone contrast using our low-order wavefront sensing and control. We summarize experimental results that quantify the performance of the testbed in terms of contrast, inner/outer working angle and bandpass, and analyze limiting factors.
\end{abstract}


\section{INTRODUCTION}

The study of worlds similar to ours where life could be present is arguably one of the most compelling science drivers for future large space observatories. The most recent Astrophysics Decadal Survey ``Pathways to Discovery in Astronomy and Astrophysics for the 2020s'' recommended a large ($\sim6\mathrm{m}$ aperture) infrared/optical/ultraviolet (LIROUV) space telescope to be developed for this purpose. This new space mission will obtain direct spectroscopic measurements of terrestrial planet atmospheres to enable the search for signatures of habitability and seek out potential molecular markers of distant biologies.  

In this context, the HiCAT testbed (High contrast imager for Complex Aperture Telescopes, \cite{ndiaye2013SPIE,ndiaye2014SPIE_hicat2,2015SPIE.9605E..0IN,2016SPIE.9904E..3CL,2017SPIE10562E..2ZL,2018SPIE10698E..53M,2018SPIE10698E..1OS}) is a dedicated coronagraphic demonstration with on-axis segmented apertures. The project is targeting system-level experiments in ambient conditions that can happen before demonstrations in vacuum. 

The project, started in 2013, has now achieved a high maturity.  In order to provide a system-like approach, the testbed includes a segmented mirror hardware simulator, a coronagraph, a low-order wavefront sensor, and other metrology capabilities (interferometric metrology, and two phase retrieval arms). The testbed has the flexibility to switch coronagraphic mode readily and in this paper we present results with a Classical Lyot Coronagraph (CLC), Apodized Pupil Lyot Coronagraph (APLC)\cite{2005ApJ...618L.161S, 2015ApJ...799..225N, 2016JATIS...2a1012Z} and Phase-Apodized-Pupil Lyot Coronagraph (PAPLC)\cite{Por2020PAPLC}.  The project work has alternated between infrastructure development and contrast performance optimization for the full system in the simpler CLC mode, while developing the other component technology in parallel (in particular the apodizer masks) for the APLC mode. 

In this paper, we give an updated overview of the testbed with recent hardware and software infrastructure updates as well as an overview of the current results.  The project has been organized in three levels of milestones, from open-loop natural ambient conditions to closed-loop control under natural and artificial drift conditions.  HiCAT has reached and exceeded the dark-hole performance for all three levels of milestones, albeit only in the monochromatic sense for now.  Broadband development was halted during the pandemic for hardware reasons and will resume by the end of 2022 with a new broadband light source. 

%%%%%%%%%%%%%%%%%%%%%%%%%%%%%%%%%%%%%%%%%%%%%%%%%%%%%%%%%%%%%%
\section{Testbed overview and project goals}

\subsection{Testbed description}

\begin{figure}[th!]
\includegraphics[width=1.0\textwidth]{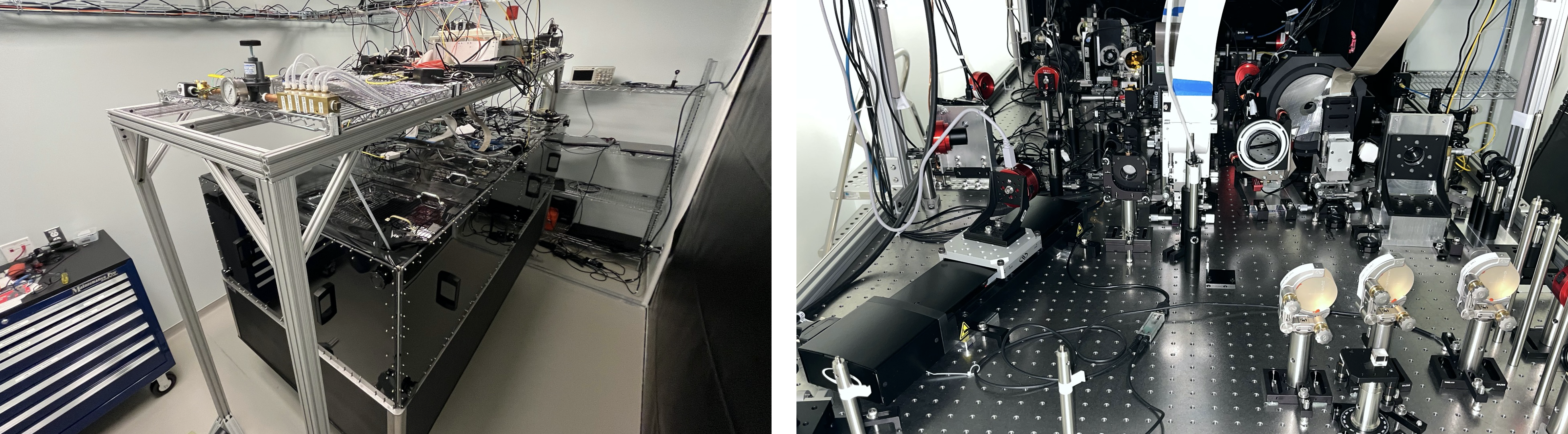}
\caption{\small HiCAT enclosure and view of the HiCAT bench itself. The three deformable mirrors can be identified by their ribbon cables. The apodizer is not in use in this picture and is replaced by a flat surrogate mirror.  }
\label{fig:picture_hicat_bench}
\end{figure}

\begin{figure}[th!]
\includegraphics[width=1.0\textwidth]{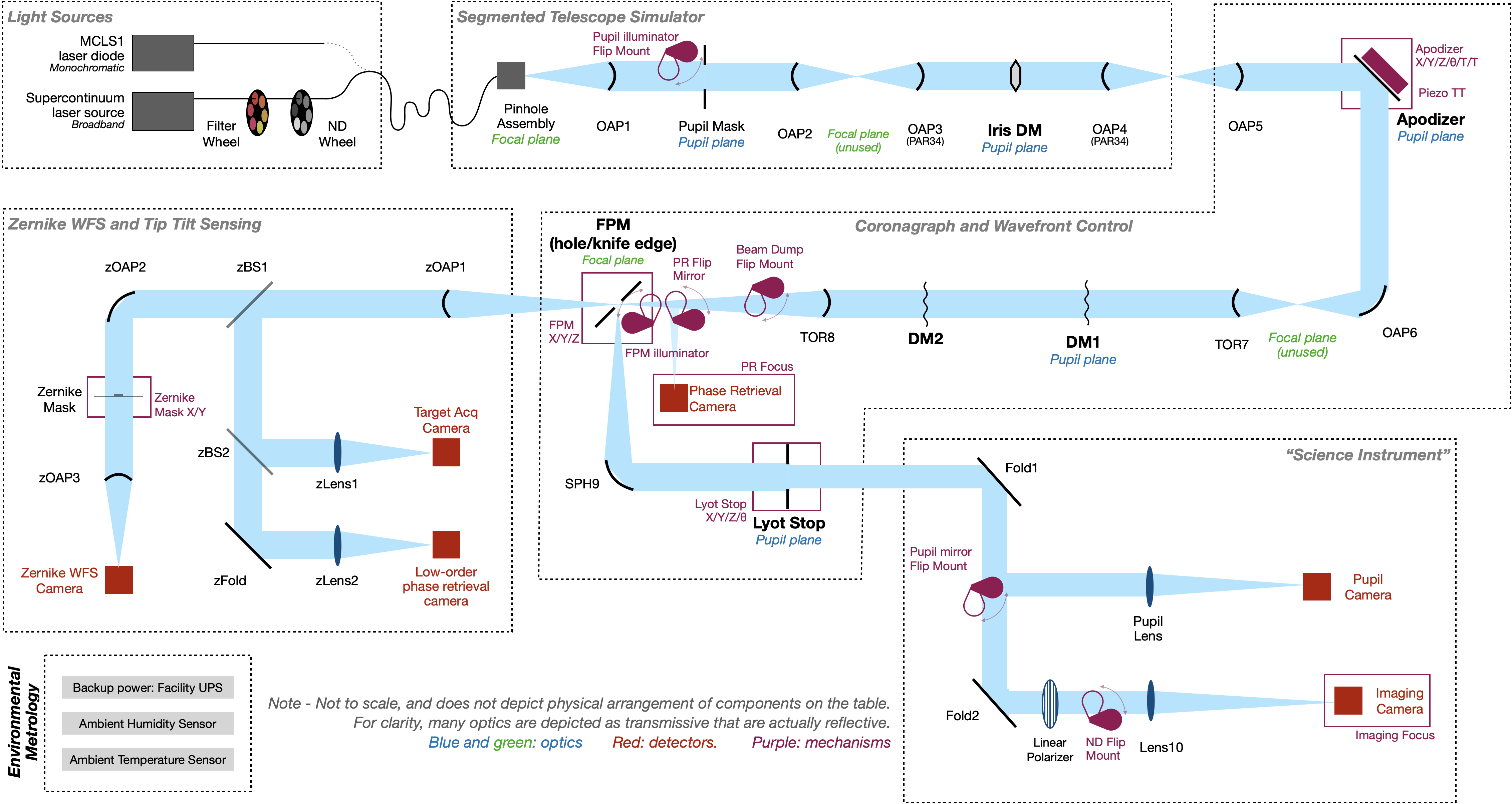}
\caption{\small HiCAT functional diagram.  Recent upgrades include the ability to swap the focal-plane mount between a circular mask for CLC/APLC mode and a knife edge for PAPLC mode. Also, a new pinhole assembly has replaced the bare single-mode fiber launch that was creating a ghost in the fiber cladding at about 3 $\lambda/D$ (similar behavior observed with two fibers). HiCAT now also has a low-order phase retrieval camera.  A new broadband laser will be installed in the fall/winter of 2022 to resume the development of broadband operations. }
\label{fig:functional_diagram}
\end{figure}

The HiCAT testbed includes a segmented aperture telescope simulator (37-segment IrisAO DM combined with an aperture stop). The entrance pupil is defined by the geometry of the 37-segment DM and is therefore not circular. The HiCAT telescope simulator is thus truly segmented with the ability to add real co-phasing wavefront errors and introduce temporal drifts for dynamical studies. 
Pictures of the HiCAT testbed are shown in Fig.~\ref{fig:picture_hicat_bench} and a functional diagram is presented in Fig.~\ref{fig:functional_diagram}.

\begin{figure}[th!]
\includegraphics[width=1.0\textwidth]{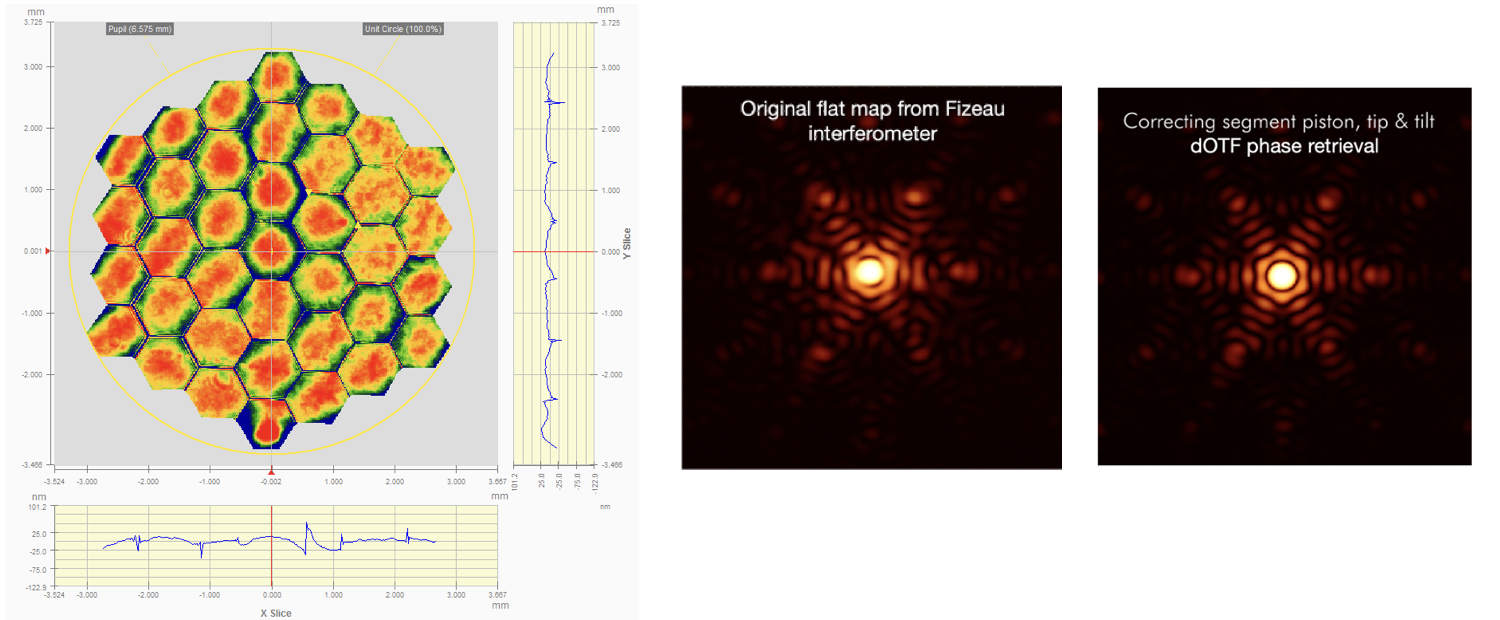}
\caption{\small  \textit{Left:} Fizeau interferometric measurement of the IrisAO surface error after manual open-loop calibration. The total surface error is $\sim9$~nm rms, and we show the corresponding PSF in the middle panel.  \textit{Right:} End-to-end calibrated PSF using dOTF\cite{Codona2013dOTF} phase retrieval with compensation of both continuous (both low-order and higher-order errors) applied to the Boston DMs, and discontinuous modes (segment-level piston/tip/tilts) on the IrisAO DM. }
\label{fig:dOTF}
\end{figure}

The segmented DM has a calibrated surface error of 9~nm rms, with very high open-loop repeatability, which makes it  suitable for the high-contrast goals of HiCAT.  Wavefront control to calibrate the wavefront further and generate dark holes (DH) is enabled by using two Boston Micromachines 952-actuator micro electro mechanical (MEMS) ``kilo-DMs". The testbed is calibrated to a final wavefront error of the order of 1~nm rms using the phase retrieval camera, which is used to measure the wavefront at a focal-plane mask (FPM) proxy location by introducing a high-quality flat mirror into the beam \cite{brady2018}. We have since then also implemented a dOTF calibration (differential Optical Transfer Function) \cite{Codona2013dOTF}, which is simpler operationally (E. Por, in prep). The dOTF calibration combines an image with a flat wavefront with an image where one of the Boston DM actuators is poked near the edge of the projected pupil to retrieve the complex amplitude of the pupil plane. During calibration, we run several iterations in closed loop to get our final corrected point-spread function (PSF), which is then used as a starting point for DH wavefront control. Figure \ref{fig:dOTF} shows images after Fizeau interferometer flattening, and after subsequent dOTF flattening.

The HiCAT coronagraph is based on a Lyot-style design and can accommodate modes including a Classical Lyot Coronagraph (CLC), Apodized Pupil Lyot Coronagraph (APLC)\cite{ndiaye2015ApJ_aplc4, ndiaye2016} and Phase-Apodized-Pupil Lyot Coronagraph (PAPLC)\cite{Por2020PAPLC}. The APLC can be implemented either using a reflective or transmissive apodizer that can be manufactured using carbon nanotubes\cite{2010SPIE.7761E..0FH}.  Unfortunately, the current state of the testbed does not allow reflective apodizers at the same time as the segmented aperture (an issue we plan to remedy in the near future by performing an optical realignment of the testbed). This issue can be mitigated in the meantime by using a transmissive apodizer, and in this paper we present results from a preliminary transmissive prototype. In CLC/APLC mode, we use a circular FPM re-used from the former ``Lyot Project'', courtesy R. Oppenheimer (American Museum of Natural History), and a transmissive Lyot stop etched in silicon and coated with carbon nanotubes. In PAPLC mode, we use a reflective knife edge that can be swapped readily using a custom kinematic mount, in conjunction with the APLC Lyot stop.  The Low-Order Wavefront sensor (LOWFS) arm includes a target acquisition camera to center the PSF precisely at the center of the FPM, and two wavefront sensors. This arm is fed with light that is rejected by the coronagraph: in CLC/APLC mode this is the light that is transmitted though the FPM hole, in PAPLC mode it is the light that misses the knife edge. Both a Zernike wavefront sensor and a Low-order Phase Retrieval (LOPR) arm are currently available and operational for closed-loop control, either by themselves or concurrently with DH algorithms.

\subsection{Project goals and milestones}

The objective is to advance high-performance coronagraph systems' technology readiness levels (TRL) for terrestrial planet direct imaging missions of the future with segmented aperture space telescopes. For such a future segmented telescope aiming to achieve a factor of $10^{-10}$ starlight suppression, the stability of the wavefront delivered from the telescope presents special challenges to the coronagraph performance. The telescope and coronagraph must be considered together and their technologies advanced as an integrated system. 

As part of the NASA Strategic Astrophysics program (Technology Demonstrations for Exoplanets Missions), the HiCAT project is organized along three formal milestones to advance system-level aspects to TRL-4 in ambient conditions\cite{TDEM-white-paper}\footnote{\url{https://exoplanets.nasa.gov/internal_resources/1186/}}. 

The DH goal for each milestone corresponds to a contrast better than $\sim 10^{-7}$ in a 360 degree dark hole extending from $\sim 4.5$ to 12 $\lambda/D$.  The milestones are organized according to three levels of complexity along the system-level demonstration axis.  The first level corresponds to a static demonstration of the dark hole.  The second level includes closed-loop LOFWS control under natural ambient drift conditions.  The third level studies the dark hole stabilization under closed loop and controlled artificial drifts that can be applied to both the segmented primary mirror segments, or the continuous DMs in the form of low-order aberrations. 

The current results of HiCAT meet the performance goals for all three milestone levels, albeit only in the monochromatic sense for now. 

\subsection{Software architecture}

\begin{figure}[th!]
\includegraphics[width=1.0\textwidth]{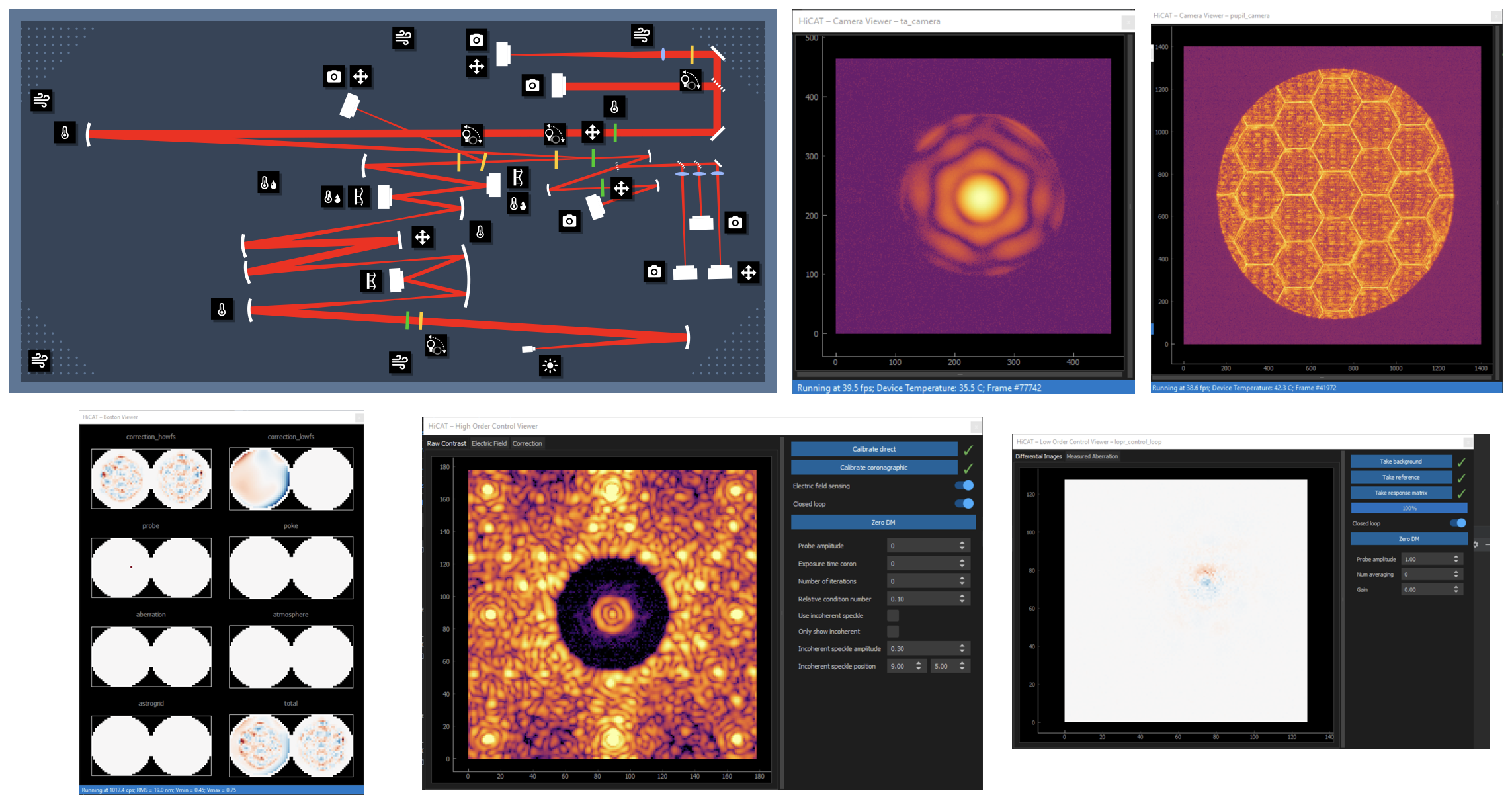}
\caption{\small A new software control infrastructure has been implemented that allows a 10x speed performance improvement, with a backend based on service-oriented architecture with low-latency inter-process communication using shared memory. From left to right and top to bottom: i) Main control panel GUI.  ii) Target-acquisition camera showing the image of the back of the FPM with the core of the PSF centered inside the mask on a logarithmic scale. iii) Pupil camera image on a logarithmic scale (here with a circular Lyot Stop in place, but non-circular entrance aperture). iv) Boston DM channels to allow for multiple independent control loops (here showing the DM surface commands for both the dark-zone control and the LOWFS correction.  v) Science camera image showing the dark zone and control GUI for sensing and control loops. vi) LOWFS control panel. 
}
\label{fig:GUI}
\end{figure}

A significant effort has been invested in developing the HiCAT software infrastructure, and in particular to enable concurrent operations with multiple closed loops (DH electric-field conjugation, or stroke minimization) at the same time as LOWFS control loops. Very recently we have upgraded our architecture (E. Por, in prep) to using a backend based on a service-oriented architecture with low-latency inter-process communication using shared memory. While the backend and communication layer are written in C++, the testbed can be operated and scripted fully from Python. An accompanying Graphical User Interface (GUI) is shown in Fig.~\ref{fig:GUI}. Camera viewers, DM viewers and control loops can all be started from inside this GUI.

The main motivation for the transition to this updated infrastructure was to enable faster operations (improvement of an order of magnitude in speed), as we had established that the testbed performance was limited by ambient drifts, especially at small inner working angles (IWA) (see Fig.~\ref{fig:CLC_dark_hole_progress} and Fig.~\ref{fig:Dark_Zone_improvement}). Furthermore, the GUI makes it easier and faster to start experiments, and observe the testbed during operation, improving the active duty cycle and usefulness of operations.

%%%%%%%%%%%%%%%%%%%%%%%%%%%%%%%%%%%%%%%%%%%%%%%%%%%%%%%%%%%%%%
\clearpage 

\section{Milestone level 1: Coronagraph component demonstration in natural conditions}
\subsection{Classical Lyot Coronagraph}
\begin{figure}[th!]
\centering
\includegraphics[width=\textwidth]{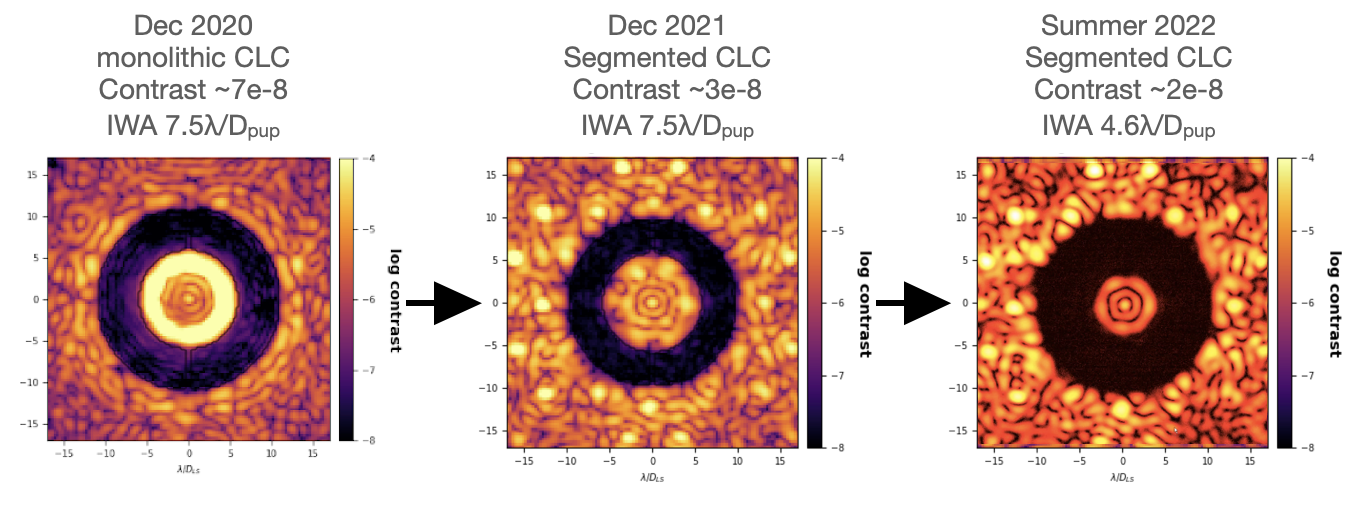}
\caption{\small Dark hole progress over the past 1.5 years, from a CLC with a circular monolithic aperture, to a truly segmented, non-circular aperture.  The contrast has improved by a factor $\sim 4$ and IWA by a factor $\sim 1.6$. }
\label{fig:CLC_dark_hole_progress}
\end{figure}

\begin{figure}[th!]
\centering
\includegraphics[width=0.8\textwidth]{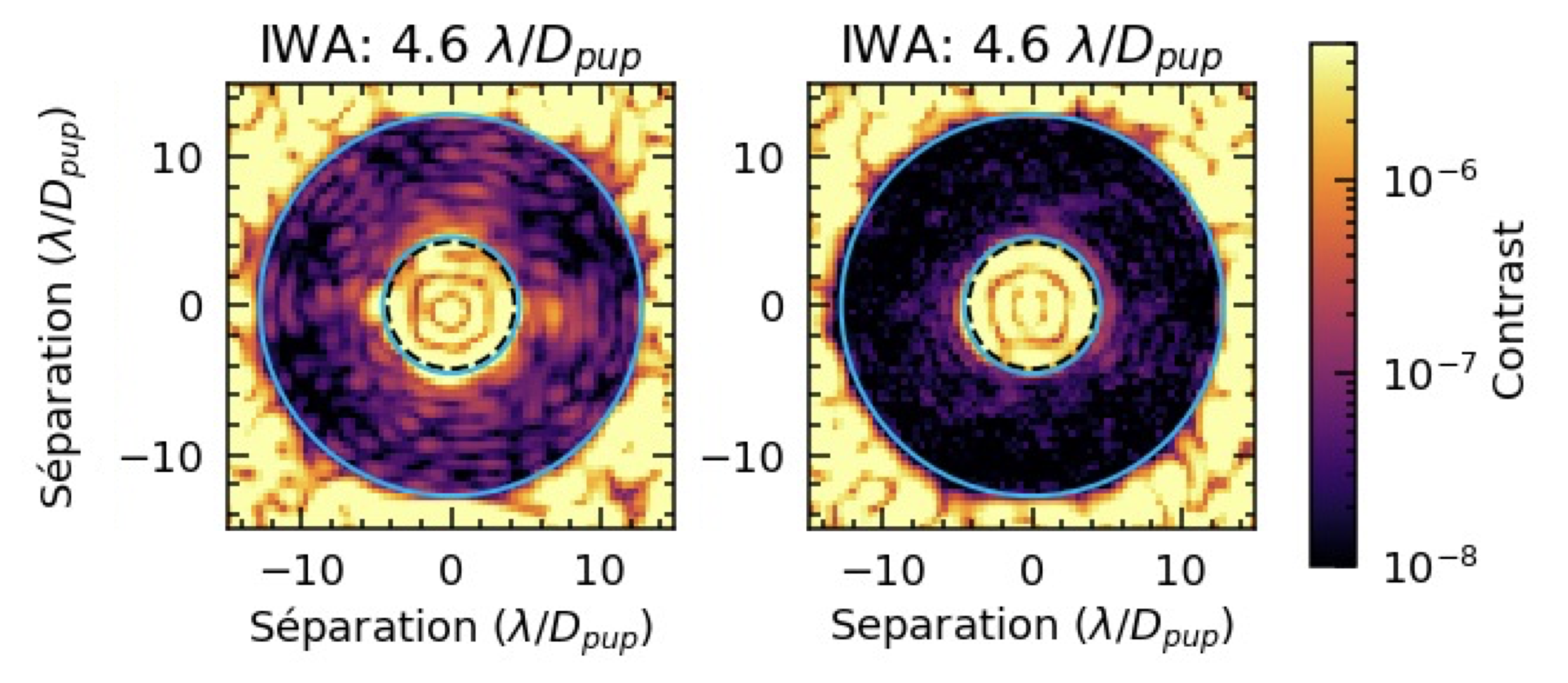}
\caption{\small Comparison between the DH performance with the previous slower infrastructure (left) and with the new faster one (right).  This is also compounded with some hardware and environment improvements.  The contrast was more limited at small IWA, and the higher speed of course plays a major role to mitigate temporal drifts.  This level of contrast is achieved on the order of a minute. Other improvements have been associated with stability (better thermal management inside the enclosure and lab HVAC repair/upgrade, as well as the installation of a pinhole. }
\label{fig:Dark_Zone_improvement}
\end{figure}

In Fig.~\ref{fig:CLC_dark_hole_progress} we show the progress made in terms of contrast and IWA over the past 1.5 years.  Initially, the testbed operated with a surrogate flat mirror in place of the IrisAO DM, with a circular aperture, and with a very large IWA ($7.5 \lambda/D_{pup}$). The testbed transitioned from this circular monolithic mode to the truly segmented non-circular aperture in the winter of 2021. A number of improvements in the used algorithm, calibration, and hardware allowed the contrast to increase by a factor 2-3x despite the more complex aperture geometry. At that point the testbed was essentially limited by stability, which was addressed by a combination of both software and hardware improvements. This allowed for a further improvement in contrast (factor $\sim 4$) and IWA (factor $\sim 1.6$). The performance impact is particularly noticeable in the IWA improvement, illustrated in Fig.~\ref{fig:Dark_Zone_improvement}, confirming that the performance was limited by the dynamical drifts and instabilities, which are now better compensated by the increased operational speed. 

\subsection{Apodized Pupil, and Phase-Apodized-Pupil Lyot Coronagraph}

\begin{figure}[th!]
\includegraphics[width=1.0\textwidth]{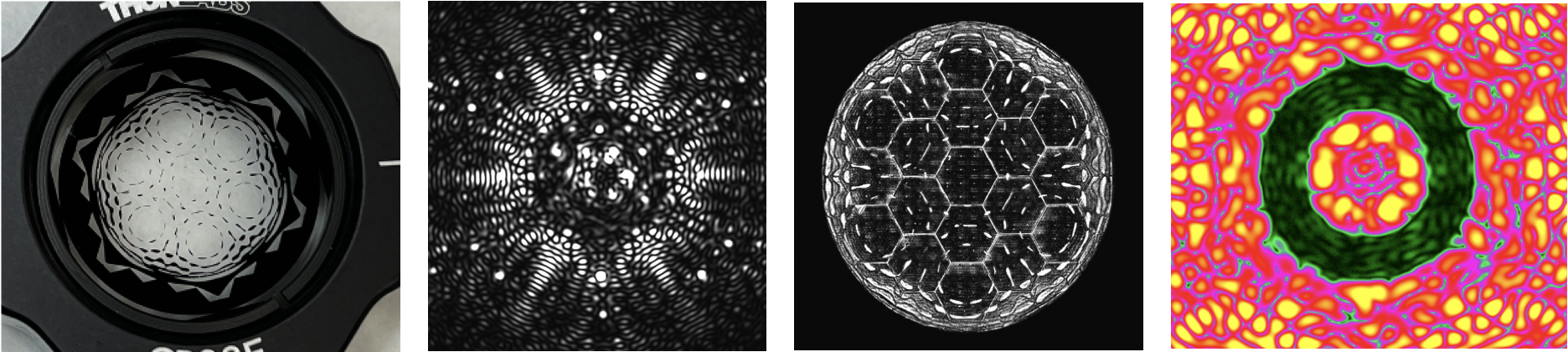}
\caption{\small First demonstration of a fully segmented aperture APLC. Because of an alignment issue that was not possible to fix in the past couple of years because of the COVID-19 pandemic, we have developed a transmissive apodizer option which allows the concurrent use of the truly segmented aperture with the apodizer. The first apodizer prototype (though impacted by some manufacturing defects) was installed. The coronagraphic PSF without wavefront control in the center left shows significant presence of wavefront error that will be calibrated in the future using phase retrieval. The contrast in the dark zone with this preliminary apodizer shown on the far right reached a 2-3x times worse value than CLC at about $6-8\times 10^{-8}$. }
\label{fig:aplc}
\end{figure}

Apodizers for HiCAT have been developed in partnership with \textit{Advanced NanoPhotonics}\footnote{\url{https://www.advancednanophotonics.com/}}\cite{2018SPIE10698E..1OS, 2022ApSS..57952250I}. After testing a number of reflective apodizers, we identified an alignment issue with the testbed that currently prevents us from using both the apodizer and the IrisAO DM at the same time.  This will be fixed eventually by some optical realignment, and at the moment we are investigating transmissive apodizers as a short-term mitigation path. Indeed, we can use the entrance pupil mask plane to insert a transmissive apodizer\cite{Ngugyen2022GPI2} (see Fig.~\ref{fig:functional_diagram}). The process for manufacturing the apodizer involves a catalyst patterned on the substrate, from which dark carbon nanotubes are grown using chemical vapor deposition at around $\sim600^{\circ}C$. The carbon nanotubes result in 100\% blocking of incoming light while minimizing reflected light (total hemispheric reflectance is 0.5\% in visible wavelengths, 0.2\% at infrared). Some of the manufacturing requirements of the apodizers are an excellent transmitted wavefront with a minimal error (we achieved $\sim 5$~nm rms) and high spectral transmission (greater than 90\% from 250-500~nm and less than 92.5\% from 500-2500~nm). The first such apodizer was received just in time for these conference proceedings, and we include here preliminary illustrations of the full segmented-aperture APLC results, shown in Fig.~\ref{fig:aplc}, with very preliminary results 2-3x worse in contrast than with the CLC ($6-8\times 10^{-8}$). However, this first prototype was impacted by a lower wavefront quality (a dOTF calibration was not performed in these results) and some coating issues, which now have been fixed in a second prototype that will be implemented on the testbed by the end of the year. 

Recently, and after a successful initial demonstration \cite{por2021SPIE}, we have upgraded the focal-plane mount to enable the PAPLC mode. This coronagraph\cite{Por2020PAPLC} uses a knife-edge FPM instead of a circular one, and does not require amplitude apodization.

The PAPLC offers a very small IWA ($\sim2\lambda/D$) which is about $\sim2-3\times$ smaller than for the CLC or APLC. This comes at the price of half a field of view.  This feature makes the two classes of concepts (PAPLC and APLC) well matched as they support distinct observing concepts for direct imaging missions. Those typically require target identification and discovery at shorter wavelengths to enable smaller IWA on the sky (where a full field of view helps) but are usually followed by characterizations over a wide range of wavelengths and in particular longer wavelengths where the PAPLC advantage is clear (a half field of view does not impact characterization since the point-source location is already known, and the smaller IWA is an obvious advantage at longer wavelengths). 

\begin{figure}[th!]
\centering
\includegraphics[width=0.9\textwidth]{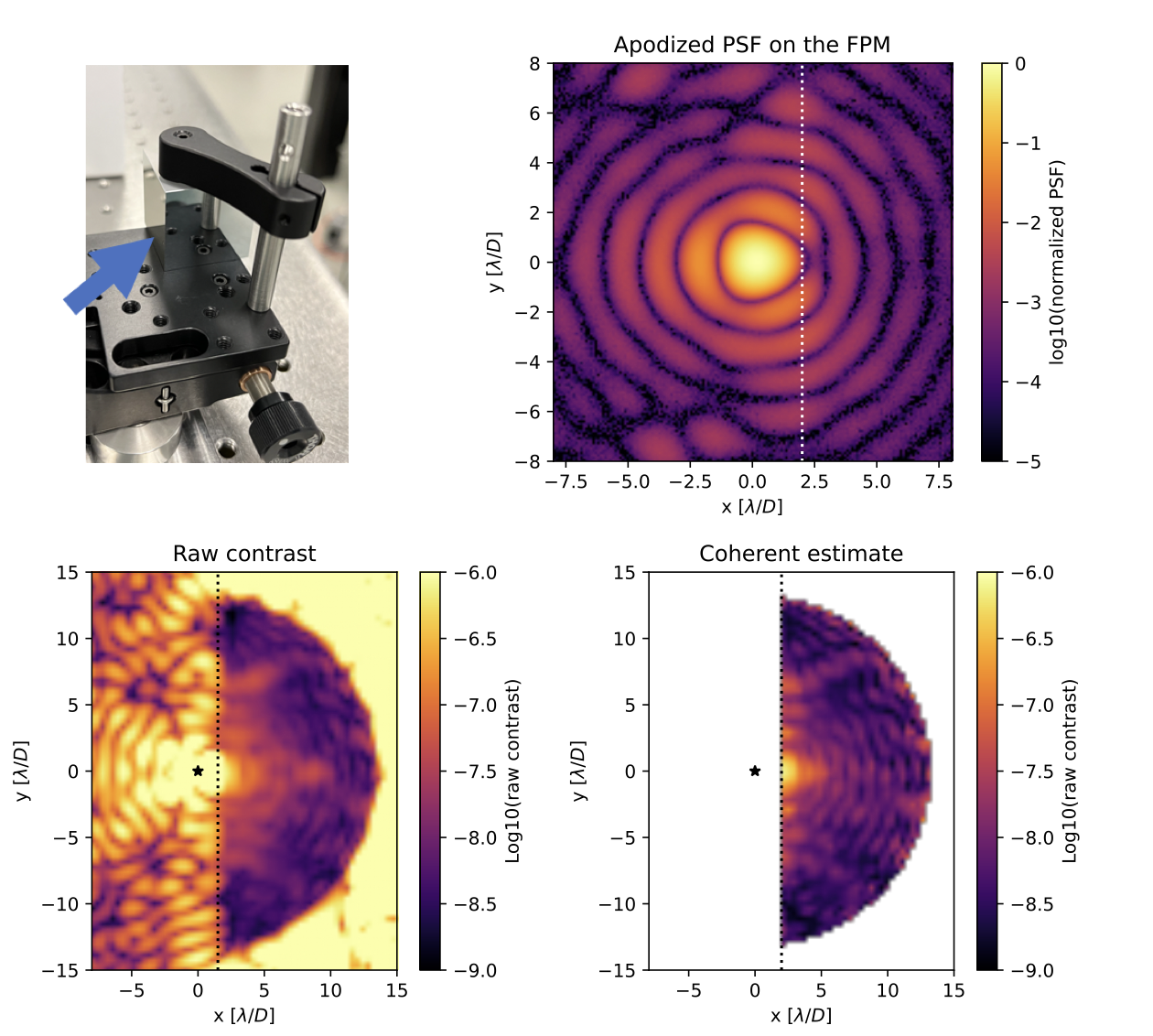}
\caption{\small The focal-plane mask mount was recently upgraded with a kinematic mount to easily swap between a circular FPM for CLC/APLC operations and a reflective knife edge (top left) for PAPLC operations. This coronagraph does not require any amplitude mask, but uses both Boston DMs to introduce phase apodization. This apodizes the PSF falling on the knife-edge FPM (top right) to suppress the stellar light in the science camera image (bottom) to reach an IWA of about $2.3\lambda/D_{pup}$ albeit in a half field of view. The raw average contrast reaches  $2\times 10^{-8}$ from 2-13 $\lambda/D_{pup}$, and $8\times 10^{-9}$ from 5-13 $\lambda/D_{pup}$.}
\label{fig:paplc}
\end{figure}

From a system-level aspect, the use of a PAPLC also offers interesting advantages since the system is a lot more sensitive at smaller angular separations. This allowed us, for example, to identify incoherent ghosts within $3.5\lambda/D$ due to unexpected propagation in the cladding of our single-mode fibers. This was confirmed to be the issue with different fibers and was finally fixed by a pinhole assembly at the beam launch. The current PAPLC performance is shown in Fig.~\ref{fig:paplc} and will be detailed in a future publication (Por et al., in prep).

%%%%%%%%%%%%%%%%%%%%%%%%%%%%%%%%%%%%%%%%%%%%%%%%%%%%%%%%%%%%%%
\clearpage
\section{Milestone level 2: Dark hole with CLOSED-LOOP CONTROL UNDER NATURAL DRIFTS}

\begin{figure}[th!]
\includegraphics[width=1.0\textwidth]{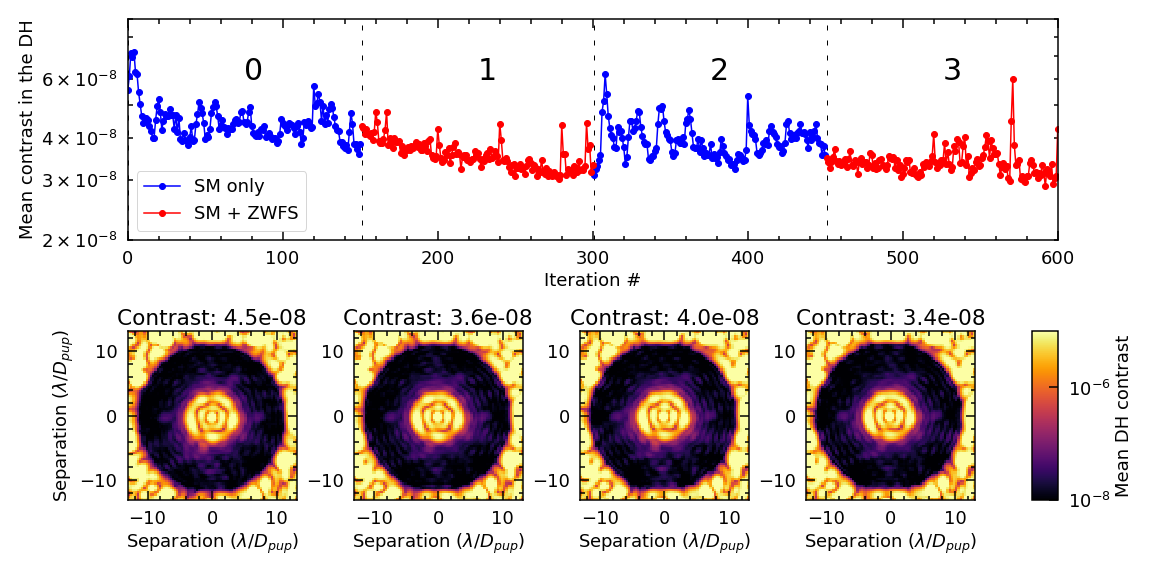}
\caption{\small Illustration of contrast performance by running simultaneous stroke minimization and Zernike control\cite{Pourcelot2022submitted} under artificially high turbulence levels (dry air purge turned on high-speed and open enclosure). This also results in a slightly worse contrast compared to the best results presented in the rest of the paper. Thanks to the increased speed of operations (stroke minimization closed loop running at 0.4\,Hz in this example over about 30 min of experiment duration, and the low-order loop at 80\,Hz), the high-order loop corrects some of the low-order drifts and the impact of the LOWFS is less significant. The contrast is nevertheless measurably lower when both loops are closed simultaneously than when the high-order loop runs by itself.    } 
\label{fig:LOWFS_simulataneous_control}
\end{figure}

In Fig.~\ref{fig:LOWFS_simulataneous_control}, we show results for simultaneous closed-loop control on both a low-speed stroke minimization high-order loop, running at $\sim0.4\mathrm{Hz}$, and a high-speed Zernike WFS low-order loop, running at $\sim80\mathrm{Hz}$. During this test, we increased our dry-air purge lines and partially opened the enclosure to let in humidity. Both of these result in higher (natural/uncontrolled) turbulence. Note that in this condition, HiCAT will not reach the same performance as shown in the previous section. While the high-order loop by itself can correct for some slow-speed drifts, Fig.~\ref{fig:LOWFS_simulataneous_control} shows that the contrast is deeper and more stable when the low-order loop is running, as seen in part 1 and 3 of the figure (red lines).  Further details and analysis of these experiments can be found in Ref.~\citenum{Pourcelot2022submitted}. 

\begin{figure}[th!]
\includegraphics[width=0.7\textwidth]{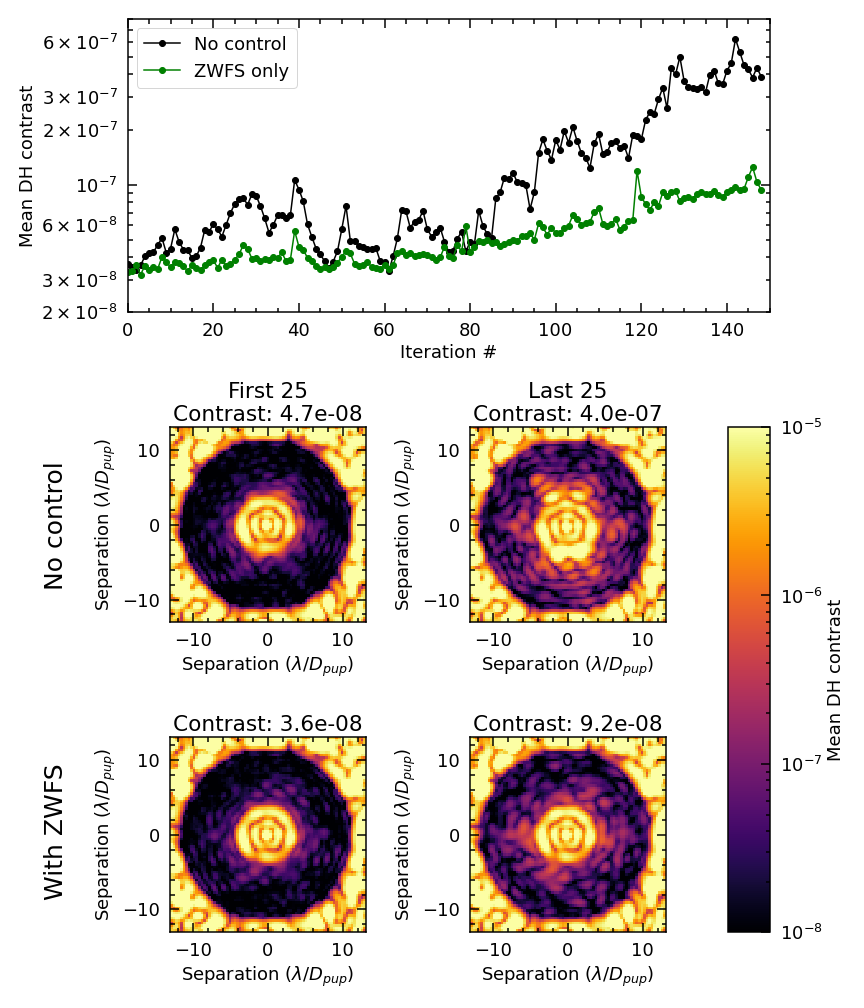}
\centering
\caption{\small Comparison of open-loop drifts without any control (i.e., simply holding the dark-hole commands on the continuous DMs) with  activated Zernike LOWFS control over 30 min with artificially high turbulence levels (dry air purge on high-speed and open enclosure). In this situation the stabilization provided by the LOWFS loop provides a 5x improvement in contrast. This operating mode corresponds to the mode of operation of the Roman Space Telescope Coronagraphic Instrument (stabilization of low-order drifts and open-loop high-order control). }
\label{fig:LOWFS_natural_drift}
\end{figure}

Natural drift results without the high-order stroke minimization loop running are shown in Fig.~\ref{fig:LOWFS_natural_drift}. Here, the Zernike WFS low-order control loop is only controlling low-order aberrations on our in-pupil Boston DM and high-orders are left static. This shows that the Zernike control loop is able to stabilize in part the source of the drifts. This experiment, combining high-order open-loop and low-order closed-loop control corresponds to the observational scenario of the upcoming Roman Space Telescope Coronagraphic Instrument (CGI) \cite{2017SPIE10400E..0DS, 2018SPIE10698E..2OS, 2019SPIE11117E..0IS}. However, it is also clear that there are drifts uncontrolled by the Zernike WFS loop that impact the contrast in the dark zone, even at timescales as short as a few minutes.

%%%%%%%%%%%%%%%%%%%%%%%%%%%%%%%%%%%%%%%%%%%%%%%%%%%%%%%%%%%%%%
\clearpage
\section{Milestone level 3: Dark hole with CLOSED-LOOP CONTROL UNDER controlled ARTIFICIAL DRIFTS}

\begin{figure}[th!]
\includegraphics[width=1.0\textwidth]{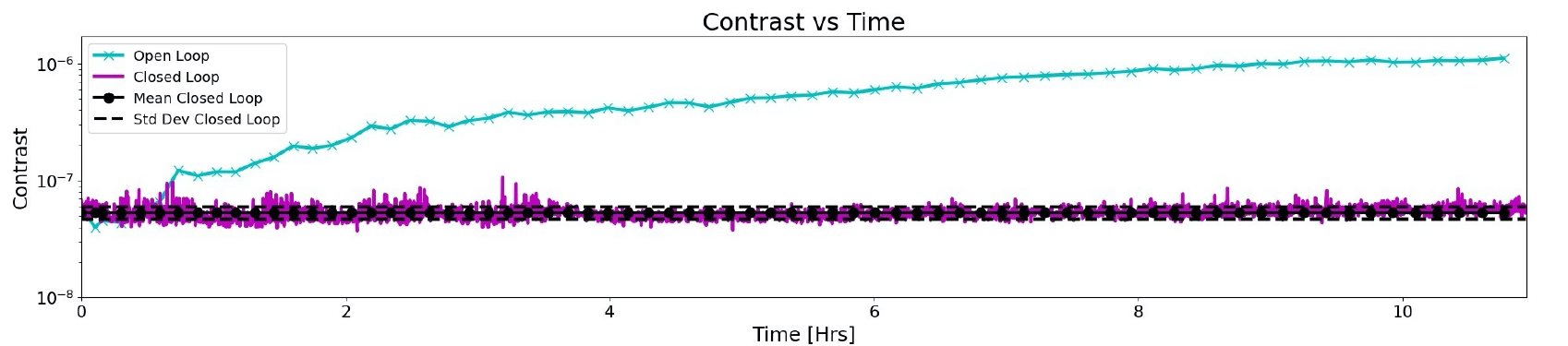}
\caption{\small Illustration of a 12 hour Dark Zone Maintenance algorithm \cite{Redmond2022SPIE} experiment where a low-SNR image is simulated using calculated Poisson statistics based on expected photon rates for a real mission (scaled to Roman CGI in this case). Artificial drifts are introduced on the Boston DM.  The closed-loop contrast (solid magenta) is maintained at $5.3\times10^{-8}$ (dotted black line) within a standard deviation of $6.4\times10^{-9}$ (dashed black line) for the duration of the experiment.  The open-loop contrast (cyan crosses) diverges to $1.1\times10^{-6}$ by the final iteration. }
\label{fig:DH_Maintenance}
\end{figure}

In addition to natural drifts, we can also add artificial drifts using our DMs, both on the continuous Boston DMs and the segmented IrisAO DM, and try to control all these drifts using our control loops. In Fig.~\ref{fig:DH_Maintenance}, we show an example of such an experiment. Artificial high-order drifts are introduced on the Boston DMs. The dark zone is controlled using the Dark Zone Maintenance algorithm\cite{2021SPIE11823E..1KR,Redmond2022SPIE}. All images fed to this algorithm are pre-processed to add photon noise, dark current, and read noise representative of the expected noise statistics for exposures in Roman CGI\cite{2017SPIE10400E..0DS, 2018SPIE10698E..2OS, 2019SPIE11117E..0IS}. When pre-processing the images, we use an exposure time of 39~s, so this 12~hr HiCAT experiment is comparable to a 35~hr Roman CGI target observation. We can see that the algorithm maintains the contrast at $\sim5.3\times10^{-8}$ throughout the 12 hour experiment (magenta line in Fig.~\ref{fig:DH_Maintenance}), even though in open loop, the contrast would have drifted to $1.1\times10^{-6}$ (cyan line in Fig.~\ref{fig:DH_Maintenance}). 

\begin{figure}[th!]
\includegraphics[width=0.8\textwidth]{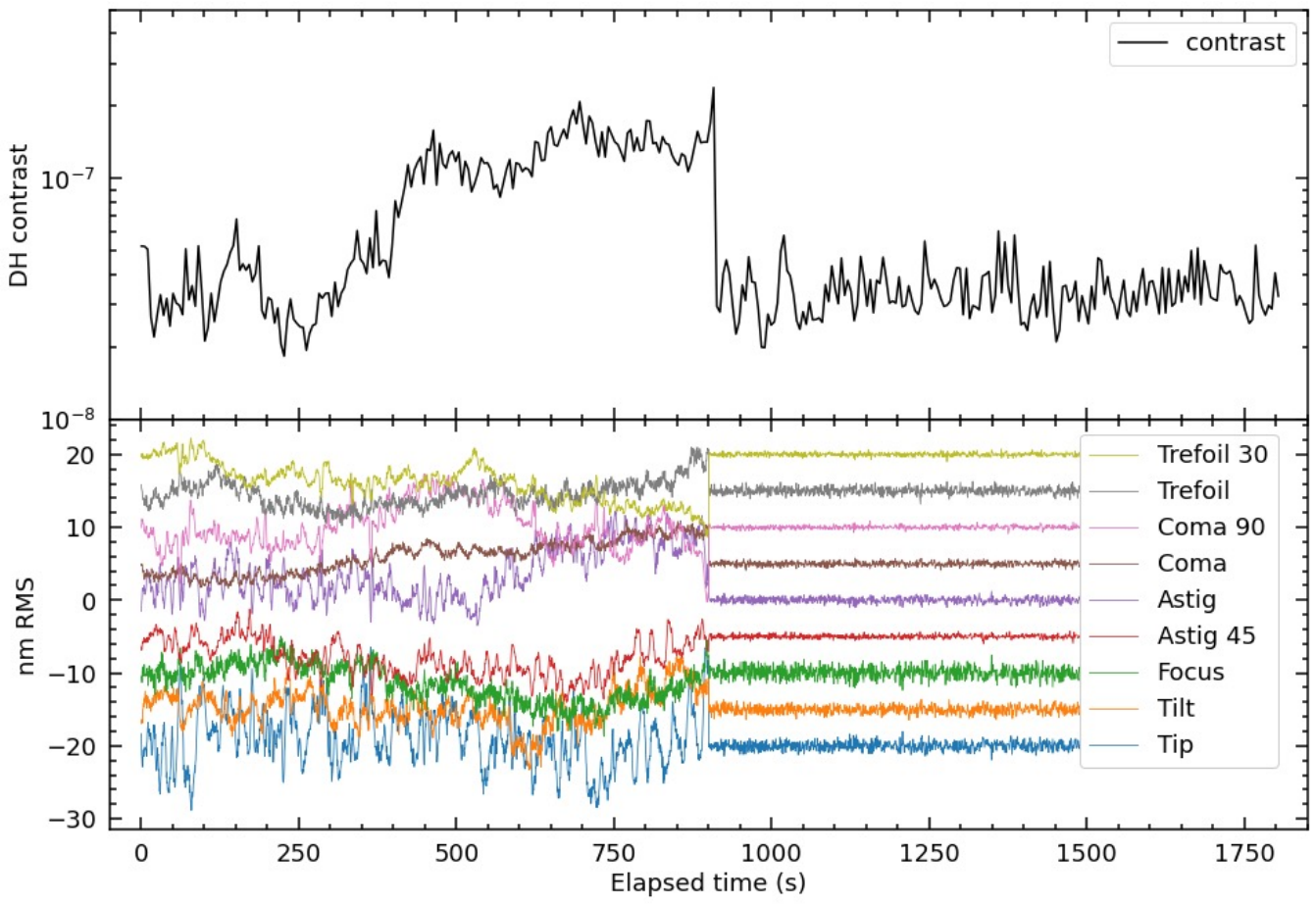}
\centering
\caption{\small Dark hole evolution with time under low-order drifts on the Boston DMs and stabilized by the Zernike LOWFS.  A random walk wavefront error is applied to the continuous Boston DMs in form of random combinations of the first ten Zernike modes. The perturbations are added every 2 seconds and represent an on-average WFE of 1.5~nm rms. The total experiment runs for 30 min and the LOWFS allows us to stabilize the contrast at about $3\times10^{-8}$, even after the open-loop drift reaches about $2\times10^{-7}$. This experiment was performed with the smaller dark zone (larger IWA) and the previoius infrastructure, and with a closed enclosure. \cite{Pourcelot2022}
}
\label{fig:LOWFS_artificial_drifts}
\end{figure}

We can also independently stabilize the low-orders aberrations using the Zernike LOWFS as in the previous section, but now under artifical drifts. The results of such an experiment is shown in Fig.~\ref{fig:LOWFS_artificial_drifts}, performed using a large IWA dark zone using the previous, slower software architecture. The Zernike control loop was started at $t=900$~sec. The first nine Zernike modes, up to trefoil and excluding piston, are nicely controlled under these artificial drifts, and the average raw contrast indeed returned to $3\times10^{-8}$, the same values as before the drifts were added. Work to combine the maintenance algorithm with low-order stabilization is underway.

%%%%%%%%%%%%%%%%%%%%%%%%%%%%%%%%%%%%%%%%%%%%%%%%%%%%%%%%%%%%%%
\clearpage
\section{Numerical and Analytical modeling}

\begin{figure}[th!]
\includegraphics[width=0.7\textwidth]{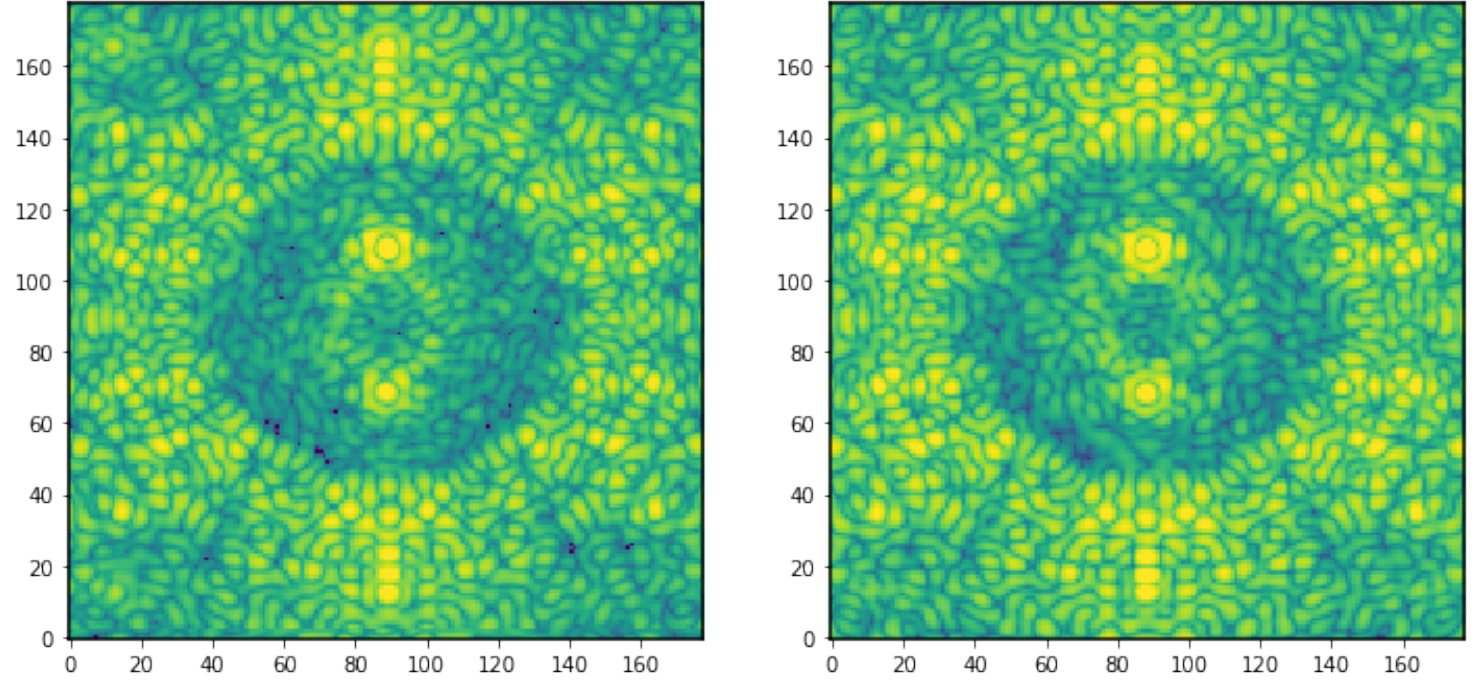}
\centering
\caption{Example output from our medium-fidelity simulator. This example is from a test of the COFFEE phase retrieval algorithm \cite{2013A&A...552A..48P,2020A&A...639A..70L} with speckle patterns applied to the DM, prior to the generation of a dark hole, in a earlier hardware mode implementing a simulated-segmented APLC (without the IrisAO DM). \textit{Left:} Actual testbed image. \textit{Right:} Simulated image for the same configuration as on the left and simulated DM settings. The overall morphology and many details are in good agreement. The main difference is that the simulated image has a lower speckle halo at wide separations (corresponding to high spatial frequencies well outside of the dark-zone control frequencies, and thus less of a priority for inclusion in the simulator thus far). We note the different intensities of the satellite spots caused by the DM actuator print-through. These asymmetries arise due to Fresnel propagation between the two DMs, which at this time were slightly misaligned in translation by about 0.2 actuator spacings. The details of this alignment and the resulting diffractive effects are reasonably well reproduced in the simulator.
\label{fig:BMM} }
\end{figure}

Simulations of the experiments performed on HiCAT allow us to calculate integration matrices needed for wavefront control, prepare new algorithms without direct testbed access and understand the limitations of the current testbed performance. HiCAT is currently modelled with a numerical simulator that implements all essential optical components (e.g., masks, DMs, cameras) but without including every single optical surface of the testbed (e.g., off-axis parabolas and lenses). This numerical model accurately reproduces images generated in all detectors on the testbed, as shown for focal-plane images on the science camera in Fig.~\ref{fig:BMM}, corresponding to an earlier testbed mode (without IrisAO, and with a reflective APLC with simulated segments). 

\begin{figure}[th!]
\includegraphics[width=1.0\textwidth]{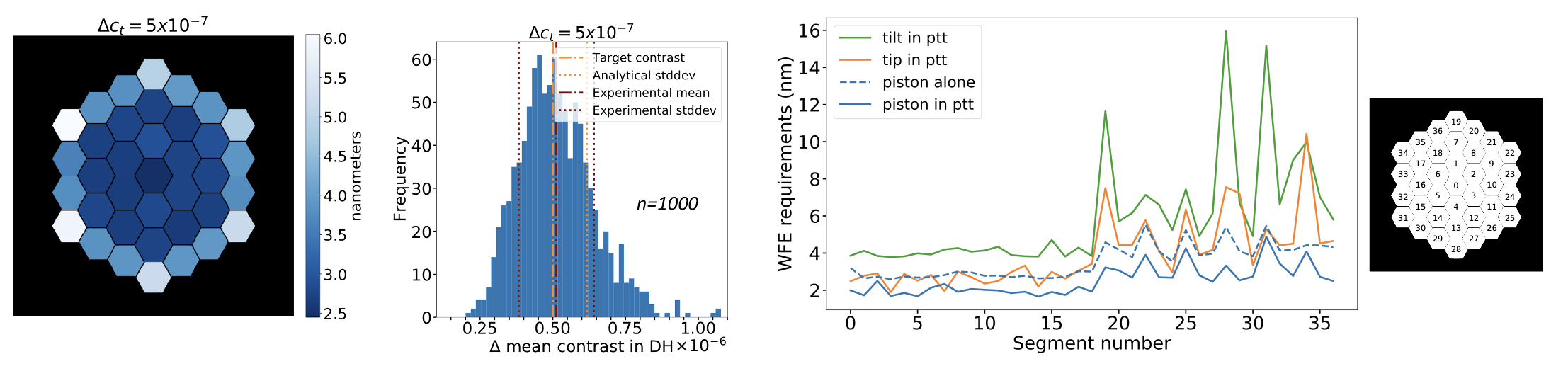}
\caption{\small \textit{Left:} Statistical segment tolerances on HiCAT (piston standard deviations, only in the un-correlated case) derived from the PASTIS sensitivity model, for a target contrast of $5\times10^{-7})$.  \textit{Center:} Validation of the tolerances on the left on hardware using a Monte-Carlo experiment. The per-segment tolerances in the left panel are used as a prescription to draw 1000 distinct aberration states on the segmented DM. The histogram represents the resulting DH contrast values. Its mean value and standard deviation coincide with the predictions of the analytical model\cite{2022A&A...658A..84L}. \textit{Right:} Generalization to multiple modes to establish the segment-level tolerances on hardware albeit at less deep target contrast ($10^{-6})$. The solid lines are the joint tolerances of piston, tip and tilt allowed on all segments. The dashed line shows the piston-only tolerances for comparison. The segment numbering on the x-axis corresponds to the segment numbering shown on the right.}
\label{fig:PASTIS}
\end{figure}

As part of the work to create comprehensive error budgets of the testbed and identify contributing factors to the performance limitation of HiCAT, analytical models like the matrix-based PASTIS sensitivity model are used.\cite{2018JATIS...4c5002L,2021JATIS...7a5004L} To quantify the impact of segment-level aberrations on the DH contrast, this model probes the contrast response to a set of localized aberration modes. By scaling the segment influences to a pre-set statistical mean contrast over many aberration realizations, we can produce tolerancing maps and requirement plots like shown in Fig.~\ref{fig:PASTIS} to establish WFE control requirements on the segmented DM. This analysis can be expanded to a wider set of local basis modes, for example through a joint requirement analysis of piston, tip and tilt aberrations on each of the segments of the IrisAO DM. In this way, we can quantify the relative contrast impact of a pertinent aberration basis set on the mean DH contrast.

This tolerancing work will benefit significantly from the improved contrast performance with the new, faster architecture to establish and validate piston/tip/tilt tolerances at higher contrast in the near future.

\section{Conclusions and perspectives}
The HiCAT project is contributing to the system-level technology maturation of a future segmented-aperture coronagraphic space mission, by demonstrating a stabilized dark zone under both natural and artificially simulated dynamic drifts in the system. This demonstration involves high-contrast dark-zone algorithms based on science camera images, along with sensing information from faster wavefront sensors, similar to what has been done for Roman CGI --- but now with the substantial added complication of segmented aberrations to sense and control. 

The HiCAT infrastructure is complete, with all hardware (integration of the Zernike LOWFS, phase retrieval and metrology paths, pinhole filter) and a software architecture to operate the tested with multiple concurrent control loops. The testbed now routinely achieves an average DH contrast of $\sim 2\times 10^{-8}$ from $4.6-13 \lambda/D_{pup}$ in CLC mode and slightly below $10^{-8}$ in the outer regions of the DH. These DHs can be obtained readily in a few minutes starting from uncontrolled DMs.  The best performance at short separations was obtained in PAPLC mode in a half dark zone with closed-loop Zernike LOWFS at a contrast of $2\times 10^{-8}$ from $2-13 \lambda/D_{pup}$, and $8\times 10^{-9}$ from $5-13 \lambda/D_{pup}$. The testbed performance is currently limited by a combination of camera performance, environment residuals from turbulence and some measurable vibrations likely due to acoustics. The resolution of the segmented IrisAO DM is appearing to be a limit for segment-drift studies, and the Boston DM electronics (14 bits) will become a limiting factor if the contrast can be improved further. 
At the moment, these results have been limited to the monochromatic configuration as we experienced hardware issues with our existing broadband light source. We have therefore focused on developing all project goals in monochromatic light. HiCAT will receive a new broadband light source by the end of 2022 and resume the broadband effort. 
 
 We have achieved all goals of the HiCAT project as defined in our SAT-TDEM white paper\cite{TDEM-white-paper} in terms of contrast and DH geometry -- albeit in monochromatic light for now, for all three milestone progression levels (DH demonstration without stabilization, DH control with LOWFS stabilization under both natural drifts and artificially controlled drifts).  

HiCAT uses three possible coronagraphic modes: CLC, APLC and PAPLC.  The development of the APLC has been delayed as we had to develop a transmissive apodizer to mitigate a pupil alignment issue preventing the use of our reflective apodizers together with the IrisAO segmented DM, until an optical realignment is performed.   The development of the APLC mode will continue with a second improved transmissive apodizer, and a new design for a reflective apodizer that is compatible with the current state of the testbed. 

The modeling component of the project has also reached its main objective by combining numerical simulations (necessary for our model-dependent control algorithms) as well as refined semi-analytical tolerancing models and validated their predictions on the testbed hardware.  The modeling effort will continue in the later phase of the project, especially to help understand the performance limits of this testbed in ambient conditions. 

\acknowledgments

This work was supported in part by the National Aeronautics and Space Administration under Grant \\
80NSSC19K0120 issued through the Strategic Astrophysics Technology/Technology Demonstration for Exoplanet Missions Program (SAT-TDEM; PI: R. Soummer). 
E.H.P. is supported by the NASA Hubble Fellowship grant \#HST-HF2-51467.001-A awarded by the Space Telescope Science Institute, which is operated by the Association of Universities for Research in Astronomy, Incorporated, under NASA contract NAS5-26555. I.L. acknowledges the support by a postdoctoral grant issued by the Centre National d'Études Spatiales (CNES) in France.  J.F.S. and L.M. acknowledge funding by the French national aerospace research center ONERA (Office National d’Etudes et Recherches Aérospatiales) and J.F.S. and M.F. by the Laboratoire d'Astrophysique de Marseille (LAM).

% References

\end{document}